# The Mid-IR Albedo of Neptune Derived from *Spitzer* Observations

Anthony Mallama

14012 Lancaster Lane

Bowie, MD, 20715, USA

anthony.mallama@gmail.com

and

Liming Li

Department of Physics

University of Houston

Houston, TX, 77004, USA

lli7@central.uh.edu

Abstract

Mid-IR albedo values of Neptune are derived from *Spitzer Space Telescope* measurements reported by Stauffer et al. (2016). The method of this derivation is described and the results indicate that the geometric albedo was about 1% or less at the time of the observations in 2016. Short-term mid-IR variability of Neptune, evidenced by the *Spitze*r observations themselves, indicates an albedo at 3.6 µm ranging from 0.2% to 0.6% with a mean of 0.4%. The corresponding albedos at 4.5 µm are 0.7%, 1.3% and 0.9%. Furthermore, the 60-year history of visible-light brightness variations, which show that Neptune was significantly fainter a few decades ago, suggests that the mid-IR albedo during that earlier period of time may have been much less than 1%. The albedo values reported here can have implications for models of Neptune's atmosphere. However, the mid-IR brightness of Neptune cannot contribute very strongly to its total albedo because the Sun emits only about 2% of its flux long-ward of 3 µm. By contrast, the Sun emits 42% of its flux at visible and near-UV wavelengths where the planet's albedo is in the tens of percents.

## 1. Introduction

The albedo of a planet is needed for deriving the body's energy balance as explained by Li et al. (2018) and in several earlier studies. The currently accepted values for Neptune were derived by Pearl and Conrath (1991) from observations obtained with the *Voyager* spacecraft. However, Li et al. have suggested that the albedo and energy balance of Neptune should be re-assessed. Their reasoning is that the Voyager-based values for Jupiter both had to be revised upward significantly based on more recent observations from the *Cassini* spacecraft and other data sets. Li et al. found that the new Jovian albedo is 47% higher and its energy balance is 37% higher. The revised energy balance has important implications for the modeling of Jupiter's interior. The same would be true for Neptune if its albedo and energy balance also need adjustments.

The purpose of this paper is to determine the mid-IR geometric albedo of Neptune based on observations from the *Spitzer* spacecraft as reported by Stauffer et al. (2016). While the Sun emits a relatively small amount of flux in the mid-IR, it will be useful to deduce the Neptunian albedo in this range of wavelengths as part of a general reassessment across all wavelengths. Furthermore, the mid-IR albedo bears upon on other aspects of Neptune including atmospheric modeling.

Section 2 of the paper briefly introduces the *Spitzer* data and illustrates that Neptune's mid-IR brightness varies strongly on rotational time-scales. Section 3 describes the  step-by-step process through which the albedo was derived. This includes converting magnitudes to flux units based on the standard star Vega, accounting for the distance to Neptune as well as its size and comparing the planet's flux to that of the Sun. Section 4 presents the albedo results, discusses the variability of Neptune and gives the conclusions.

2. The *Spitzer* data

Stauffer et al. (2016) reported brightness measurements of Neptune obtained at 3.6- and 4.5-µm in channels 1 and 2 of the IRAC instrument onboard the *Spitzer Space Telescope.* The observations were obtained during 2016 February 21-23 and the set of data from each channel covered about one rotation of the planet. The authors principally analyzed the variability of the brightness and compared it with other IR data from the *WISE* spacecraft as well as visible light measurements from the *Hubble Space Telescope* and the *Kepler/K2* mission. They concluded that the wavelength-dependent brightness variations are due to a cloud deck with great contrast situated higher in the atmosphere than much of its methane opacity.

The brightness measurements are given in magnitude units in tables 1 and 2 of Stauffer et al. Those data were downloaded from the web-site of the paper for this study. The light-curves are shown here as Fig. 1.

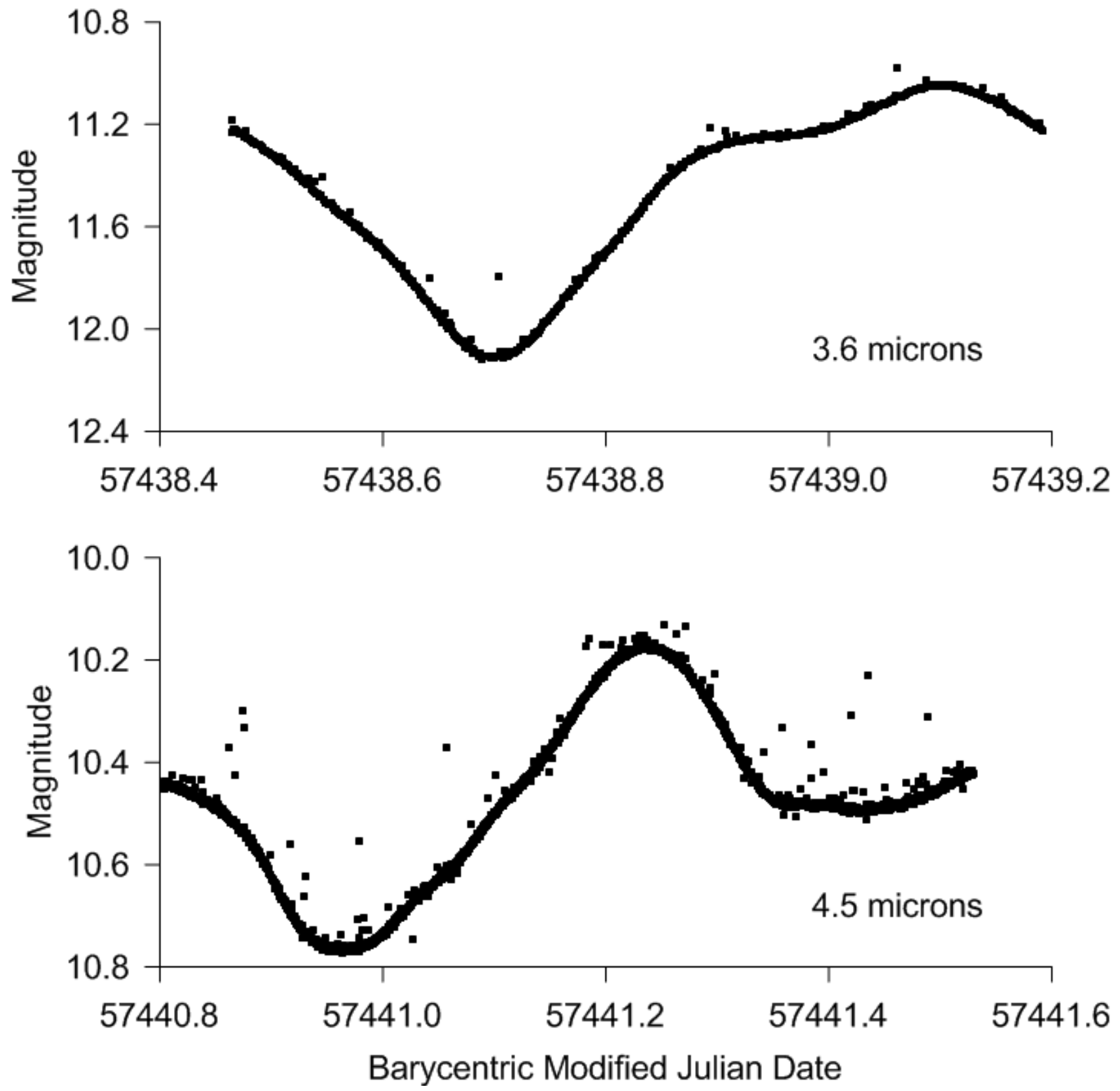

*Fig. 1. Spitzer light-curves. The magnitude data from tables 1 and 2 of Stauffer et al. show the variable brightness of Neptune over approximately one rotation of the planet. Two outlying magnitudes were removed from the 4.5 micron data which contained more than 2,000 points. The remaining data were then plotted and analyzed.*

## 3. Method of deriving albedos

The general outline for computing Neptunian albedos is as follows. Magnitudes were converted to flux units using the zero points at 3.6- and 4.5-µm which are given by Stauffer et al. (2016) in Jansky (Jy) units. Those Jy fluxes, in turn, were converted to flux units per wavelength unit rather than per frequency unit in order to compare with a table of solar flux which is given in wavelength units. Neptune's distances from the Sun and from the *Spitzer* spacecraft were factored in to determine its flux at one AU, again for use with the solar flux. The ratio of luminosity between Neptune at the Sun was computed. Finally, the radius of Neptune was factored in to give the geometric albedo. These steps are elaborated below and are shown with their numerical values in the appendix to this paper. The appendix contains separate sections for 3.6- and 4.5-µm values.

### 3.1 Converting flux units

Several data sets with different units are need for these computations. The conversion factor between ergs/s/cm^2/A and W/m^2/um is derived here and the result is used in section 3.3. There is also a worked example in the appendix where a sample solar flux value in W/m^2/µm from the E490 data set of the National Renewable Resources Data Center ( http://rredc.nrel.gov/solar/spectra/am0/) is compared to that from the *Hubble Space Telescope* solar flux (ftp://ftp.stsci.edu/cdbs/current_calspec) in ergs/s/cm^2/A at the same wavelength.

### 3.2 Converting from frequency to wavelength units

The magnitudes given by Stauffer et al. are referenced to the flux of the standard star Vega in Jy units which are frequency-based. Therefore it is necessary to derive the ratio between spacing of frequency (Hz) and that of wavelength (µm) units in the 3.6- and 4.5-µm *Spitzer* channels. This allows conversion between frequency-based and wavelength-based fluxes.

### 3.3 Vega flux

The flux of Vega as given by Stauffer et al. at 3.6- and 4.5-µm is provided here. It is compared to the flux of Vega from the *Hubble* STIS (ftp://ftp.stsci.edu/cdbs/current_calspec) instrument for validation. The agreement is 3% at 3.6-µm and 1% at 4.5-µm.

### 3.4 Neptune magnitudes

The 3.6- and 4.5-µm data sets contain 623 and 2016 magnitudes, respectively. This section summarizes their statistics, namely the minimum, mean and maximum values.

### 3.5 Neptune fluxes

The fluxes of Neptune (maximum, mean and minimum) are computed based upon the standard star Vega in two ways. The first is from frequency units deriving from Stauffer's flux and the second is in wavelengths units deriving from STIS's Vega flux. The results agree at the same level as those in section 3.3.

### 3.6 Distance factor

A correction factor is needed to compute the brightness of Neptune at a distance of one AU from the *Spitzer* spacecraft and from the Sun. This will be needed in the next section. According to HORIZONS, the Sun-to-Neptune distance for the observing period of 2016 February 21/23 was 29.96 AU. Meanwhile the distance from spacecraft-to-Neptune was 29.76 AU. The distance factor to be applied to flux is the product of the squares of the Sun-to-Neptune and spacecraft-to-Neptune distances.

### 3.7 Neptune fluxes at 1 AU

The distance factor computed in section 3.6 is applied to the Neptune fluxes from section 3.5 to give the maximum, mean and minimum flux values at 1 AU. The fluxes are based on the fluxes of Vega given in Jy units by Stauffer et al. rather than the HST STIS fluxes of Vega.

### 3.8 Solar flux at 1 AU

The values of solar flux from the E490 model of the National Renewable Resources Data Center at 3.6- and 4.5-µm are given.

### 3.9 Luminosity ratio

The ratio of Neptune's flux to solar flux at 1 AU are determined by dividing the values from section 3.7 by those from 3.8.

### 3.10 Neptune size factor

The radius of Neptune is divided by the distance of 1 AU. The square of that result becomes the factor to be applied to the luminosity ratio. (To be exact, it is the square of the sine of the angle but since the angle is very small the square of the result of the division itself suffices.)

### 3.11 Neptune albedo

The luminosity ratios are divided by the Neptune size factor to give the albedo values.

## 4. Albedo results, discussions and conclusions

### 4.1 Albedo results

The maximum, mean and minimum albedos from *Spitzer* 3.6- and 4.5-µm observations are listed in Table 1. The only value which exceeds 1% corresponds to the maximum brightness at 4.5-µm. The other albedo results are all 0.9% or less. These very low values compare fairly well with the mid-IR albedos of Jupiter shown in figure 2 of Li et al. (2018).

Table 1. Neptune mid-IR albedos

| | Wavelength | |
|---|---|---|
| Luminosity | 3.6 microns | 4.5 microns |
| Maximum | 0.6 % ± 0.3% | 1.3 % ± 0.6% |
| Mean | 0.4 % ± 0.1% | 0.9 % ± 0.2% |
| Minimum | 0.2 % ± 0.1% | 0.7 % ± 0.4% |

### 4.2 Variability

Neptune was observed to vary by about a factor of 3 at 3.6-µm and a factor of 2 at 4.5-µm. Furthermore, these observations only covered about one revolution of the planet. Therefore it is very likely that the full range of brightness variation is even larger and it is also possible that the computed mean does not approximate the long-term mean. Thus, variability appears to be the most significant uncertainty in deriving an albedo values for the timeframe of the observations. We estimate an uncertainty of 25% for the mean values and 50% for the extreme values as indicated in Table 1.

Significant long-term variability is illustrated in Fig.2 which plots visible light magnitudes in the V- and y-bands (0.549 and 0.551 µm, respectively) for Neptune during the past 60 years. The planet brightened by about 0.1 magnitude or 10% in luminosity during the period from 1980 to 2000. There are other smaller variations before and after that brightening period as well.

The 10% variation in visible light may indicate much larger changes in the mid-IR. Stauffer et al. (2016) compare the variations of more than 100% that they observed in the mid-IR to those reported by Simon et al. (2016) at shorter wavelengths. Simon et al. found only a 10% brightness change in *Hubble* data at 0.845 µm, and an even smaller variation of 2% in *Kepler* data (0.400 – 0.865 µm). While these observations at different wavelengths are not contemporaneous they are very suggestive of strongly increasing variability with wavelength. That trend agrees with models of clouds and methane in the Neptunian atmosphere. Therefore the 10% visible light variation over the past 60 years may indicate a mid-IR brightness variation which is much greater than the 100% recorded by *Spitzer*. In particular, the mid-IR brightness may have been very much smaller prior to 1980 than at present. While it is impossible to validate that speculation about the albedo in the past, continued monitoring in the future could confirm or deny it.

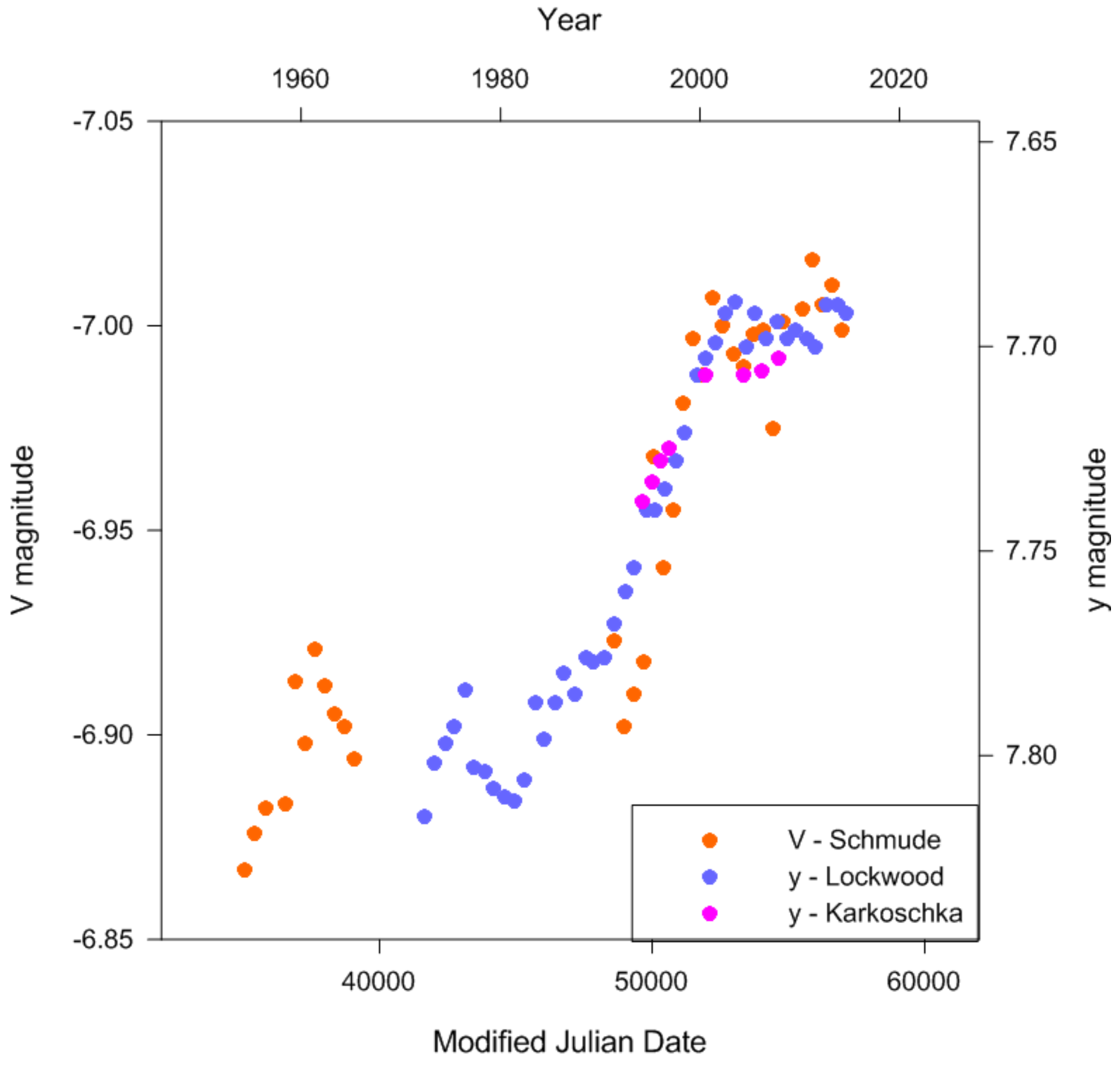

*Fig. 2. The visible light-curve of Neptune over the past 60 years based on figure 8 of Schmude et al. (2016) which includes data from Lockwood's web page at* [*http://www2.lowell.edu/users/wes/U_N_lcurves.pdf*](http://www2.lowell.edu/users/wes/U_N_lcurves.pdf) *and from* [*Karkoschka (2011)*](). 

4.3 Uncertainties

The major uncertainty in the albedo values is due to Neptune's variability as discussed in section 4.2. There is also a minor uncertainty indicated in the appendix because the fluxes of Vega taken from the *Hubble* STIS table and those indicated by Stauffer et al. (2016) do not agree exactly. Additionally, Stauffer et al. also mention an uncertainty in the flux values of Neptune due to measuring an extended object with a finite-sized photometric aperture. A correction for aperture size of about 5% was applied, however it corresponded to a source having a stellar point spread function. The apparent diameter of Neptune was 2 arc-seconds while the diameter of the aperture was 12 arc-seconds so some small amount of scattered light (besides the 5%) fell outside of the aperture. The authors consulted the Infrared Science Archive and they replied that the light lose for the extended disk of Neptune would be ‘not very different’ from that for a stellar source. A final small uncertainty is that the phase angle of Neptune (Sun-to-Neptune-to-*Spitzer*) has not been taken into account in order to compute the planet’s geometric albedo. However, that angle can only be as large as 1.9 degrees and the effect on brightness is practically nil. In any case, these uncertainties are insignificant in comparison to those stemming from the large mid-IR brightness variations considered above.

## 4.4 Contribution to total albedo

Whether or not the albedos in Table 1 represent the long-term, the mid-IR brightness of Neptune can only contribute a very small amount to the flux to the planet integrated over all wavelengths. This follows from the small percentage of solar flux that is emitted in the mid-IR. Wehrli (1985) lists the cumulative solar flux by wavelength in the table provided by the National Renewable Resources Data Center ( http://rredc.nrel.gov/solar/spectra/am0/) . The cumulative flux value at 3 μm is given as 1341 W/m^2 while the value out to 10 μm (which includes practically all of the solar flux) is 1367. Thus only about 2% of the solar flux is emitted long-ward of 3 μm. [The total solar flux reported more recently (e.g., Kopp and Lean, 2011) revises that total downward slightly but the percentage above 3 μm is still about the same.] Meanwhile the geometric albedo in the visible and near-UV portion of the solar spectrum (0.35-0.70 μm) is in the tens of percents (see the values in the Sloan u’, g’ and r’ bands in table 7 of Mallama et al. 2017) and the Sun emits 42% of its flux there. So, that shorter wavelength contribution to Neptune’s brightness overwhelms the contribution from the mid-IR.

## 4.5 Conclusions

Our conclusions are as follows. (1) Albedos derived from measurements from *Spitzer* indicate that the geometric albedo of Neptune in the mid-IR was about 1% or less at the time of the observations. (2) The 60-year history of visible-light brightness variations, which shows that Neptune was significantly fainter prior to 2000, may indicate that the mid-IR albedo during that period of time was much less than 1%. (3) The mid-IR brightness of Neptune cannot contribute very strongly to its total albedo because the Sun emits only about 2% of its flux long-ward of 3 μm.

## Acknowledgments

Amy Simon helpfully answered questions about *Spitzer* data files. The Infrared Science Archive provided the distance from *Spitzer* to Neptune and information about light lost in aperture photometry.

References

Karkoschka, E. 2011. Neptune's cloud and haze variations 1994–2008 from 500 HST–WFPC2 images. Icarus 215, 759-773.

Kopp, G. and Lean, J.L. 2011. A new, lower value of total solar irradiance: Evidence and climate significance. Geophys. Res. Let. 38, L01706.

Li, L., Jiang, X., West, R.A., Gierasch, P.J., Perez-Hoyos, S., Sanchez-Lavega, A., Fletcher, L.N., Fortney, J.J., Knowles, B., Porco, C.C., Baines, K.H., Fry, P.M., Mallama, A., Achterberg, R.K., Simon, A.A., Nixon, C.A., Orton, G.S., Dyudina, U.A., Edwald, S.P. 2018. Less absorbed solar energy and more internal heat for Jupiter. Submitted to Nature Communications.

Mallama, A., Krobusek, B. and Pavlov, H. 2017. Comprehensive wide-band magnitudes and albedos for the planets, with applications to exo-planets and Planet Nine. Icarus 282, 19-33. (Also posted here.)

Pearl, J.C. and Conrath, B.J. 1991. The albedo, effective temperature and energy balance of Neptune as determined by Voyager data. J. Geophys. Res. 96, 18921-18930.

Schmude, R.W., Baker, R.E., Fox, J., Krobusek, B.A., Pavlov, H. and Mallama, A. 2016. The secular and rotational brightness variations of Neptune.

Simon, A.A.., Rowe, J.F., Gaulme, P., Hammel, H.B., Casewell, S.L., Fortney, J.J., Gizis, J.E., Lissauer, J.J., Morales-Juberias, R., Orton, G.S., Wong, M.H., Marley, M.S. 2016. Neptune's dynamic atmosphere from Kepler K2 observations: implications for brown dwarf light curve analyses. Astrophys. J. 817:162.

Stauffer, J., Marley, M.S., Gizis, J.E., Rebull, L., Carey, S.J., Krick, J., Ingalls, J.G., Lowrance, P., Glaccum, W., Kirkpatrick, J.D., Simon, A.A., Wong, M.H. 2016. Spitzer Space Telescope mid-IR light curves of Neptune. Astron. J. 152:142.

Wehrli, C. 1985. Extraterrestrial solar spectrum, Publication no. 615, Physikalisch-Meteorologisches Observatorium + World Radiation Center (PMO/WRC) Davos Dorf, Switzerland.

Appendix – Computation of the albedo values

Neptune albedos computed from flux values in tables 1 and 2 of Stauffer et al.
" Spitzer Space Telescope Mid-IR Light Curves of Neptune". 2016. AJ. 152: 142.
doi:10.3847/0004-6256/152/5/142

**Result: albedo ranges from 0.2 to 0.6% at 3.6 um and 0.7 to 1.3% at 4.5 um.**

From Stauffer "We converted aperture fluxes to magnitudes using the inband flux densities of Vega: 278 Jy (3.6 µm) and 180 Jy (4.5 µm)."
From Wiki: The Jansky unit "... is equivalent to 10^-26 watts per square metre per hertz."

Speed of light:

| | | |
|---|---|---|
| C | 3.00E+08 | m/s |
| | 3.00E+14 | um/s |

Steps:
1. Converting flux units (compare E490 and HST STIS solar fluxes)
2. Convert Hertz to microns
3. Vega flux
4. Neptune magnitudes and range
5. Neptune flux (W/m^2/um)
6. Distance factor
7.Neptune flux (W/m^2/um,  1 AU from Earth and Sun, from Stauffer Vega Jy)
8. Solar flux at 1 AU
9. Luminosity ratio (Neptune to Sun at 1 AU)
10. Neptune size factor
11. Neptune albedo

**<------------------------------ 3.6 microns ------------------------------>**

**1. Converting flux units (compare E490 and HST STIS solar fluxes)**

| | | |
|---|---|---|
| 1 W = | 1.00E+07 | ergs/s |
| 1 m^2 = | 1.00E+04 | cm^2 |
| um = | 1.00E+04 | A |
| nm = | 1.00E+01 | A |

So, 1 W/m^2/um = 0.10 ergs/s/cm^2/A

Example:

| um | W/m^2/um | Source |
|---|---|---|

| | | | |
|---|---|---|---|
| E490 | 1.0 | 749.9 | http://rredc.nrel.gov/solar/spectra/am0/A |
| | Angstrom | ergs/cm^2/s/A | |
| HST STIS | 10000.0 | 73.2 | ftp://ftp.stsci.edu/cdbs/current_calspec |

Note that the STIS value is almost exactly 10 times smaller as expected

**2. Converting Hertz to microns**

| | | |
|---|---|---|
| Hz to um | nu = c / lambda | |
| | microns | Hz |
| | 3.5995 | 8.33E+13 |
| | 3.6005 | 8.33E+13 |
| diff | 0.001 | 2.31E+10 |
| diff / diff | | 2.31E+13 |

**3. Vega flux**

| | | | |
|---|---|---|---|
| Stauffer | 278 | Jy | (Jy = 10^-26 W/m^2/Hz) |
| | 2.78E-24 | W/m^2/Hz | |
| | 6.43E-11 | W/m^2/um | |

STIS (for checking conversion from Jy to W/m^2/Hz and general validation)

| | | |
|---|---|---|
| Angstroms | | |
| 36000 | 6.23E-12 | ergs/s/cm^2/A (from table) |
| | 6.23E-11 | W/m^2/um |

Stauffer / STIS = 1.03 OK

**4. Neptune magnitudes and range**

From statistics below

| | |
|---|---|
| 10.9784 | Min |
| 11.4656 | Mean |
| 12.1217 | Max |

**5. Neptune flux (W/m^2/um)**

Vega zero-point from Stauffer in Jy:

| | |
|---|---|
| 2.61E-15 | Max |
| 1.67E-15 | Mean |
| 9.11E-16 | Min |

Vega from HST STIS in ergs/s/cm^2/A for checking

| | | | |
|---|---|---|---|
| | | Ratio Stauffer to STIS | |
| 2.53E-15 | Max | 1.03 | OK |
| 1.62E-15 | Mean | 1.03 | OK |
| 8.83E-16 | Min | 1.03 | OK |

**6. Distance factor**

Date 2016-Feb-21/23

| | | |
|---|---|---|
| Solar | 29.96 | heliocentric distance from HORIZONS |
| Spitzer | 29.76 | based on result from HORIZONS (target '@-79') |
| | | |
| Factor | 794968 | Spitzer distance above |

**7.Neptune flux at 1 AU from the Sun and Spitzer (W/m^2/um, from Stauffer Vega Jy)**

| | |
|---|---|
| 2.01E-09 | Max |
| 1.28E-09 | Mean |
| 7.02E-10 | Min |

**8. Solar flux at 1 AU**

E-490

13.07 W/m^2/um

(Wehrli gives 0.013 W/m^2/nm at 3575.0 nm)

**9. Luminosity ratio (Neptune to Sun at 1 AU)**

| | |
|---|---|
| 1.54E-10 | Max |
| 9.83E-11 | Mean |
| 5.37E-11 | Min |

**10. Neptune size factor**

| | |
|---|---|
| 24622 | radius, km |
| 1.50E+08 | AU, km |
| 2.71E-08 | sin^2( r / AU ) |

**11. Neptune albedo**

| | | | |
|---|---|---|---|
| 0.0057 | Max | 0.6 | % |
| 0.0036 | Mean | 0.4 | % |
| 0.0020 | Min | 0.2 | % |

**Statistics of the observations below**

| | BMJD | Mag | Mag uncert. |
|---|---|---|---|
| Mean | 57438.82741 | 11.4656 | 0.0025 |
| St. Dev. | 0.21059 | 0.3329 | 0.0015 |
| SD Mean | 0.00844 | 0.0133 | 0.0001 |
| Count | 623 | 623 | 623 |
| Min | 57438.46455 | 10.9784 | 0.0015 |
| Max | 57439.19128 | 12.1217 | 0.0171 |
| Max - Min | 0.72673 | 1.1433 | 0.0156 |

[Note: In the original spreadsheet the individual 3.6 micron observations follow at this point.]

**<------------------------------ 4.5 microns ------------------------------>**

**1. Converting flux units (compare E490 and HST STIS solar fluxes)**

Same as for 3.6 microns

Hz to um nu = c / lambda

| | microns | Hz |
|---|---|---|
| | 4.4995 | 6.66E+13 |
| | 4.5005 | 6.66E+13 |
| diff | 0.001 | 1.48E+10 |
| diff / diff | | 1.48E+13 |

**3. Vega flux**

| | | | |
|---|---|---|---|
| Stauffer | 180 | Jy | (Jy = 10^-26 W/m^2/Hz) |
| | 1.80E-24 | W/m^2/Hz | |
| | 2.66E-11 | W/m^2/um | |

STIS (for checking conversion from Jy to W/m^2/Hz and general validation)

| Angstroms | | |
|---|---|---|
| 44960 | 2.65E-12 | ergs/s/cm^2/A (from table) |
| 45010 | 2.64E-12 | ergs/s/cm^2/A (from table) |
| 45000 | 2.64E-12 | ergs/s/cm^2/A (interpolated) |
| | 2.64E-11 | W/m^2/um |

Stauffer / STIS = 1.01 OK

**4. Neptune magnitudes and range**

From statistics below (note - removed 2 outlying data points as indicated in the listing below)

| | |
|---|---|
| 10.1324 | Min |
| 10.4770 | Mean |
| 10.7736 | Max |

**5. Neptune flux (W/m^2/um)**

Vega zero-point from Stauffer in Jy:

| | |
|---|---|
| 2.36E-15 | Max |
| 1.72E-15 | Mean |
| 1.31E-15 | Min |

Vega from HST STIS in ergs/s/cm^2/A for checking

| | | Ratio Stauffer to STIS | |
|---|---|---|---|
| 2.34E-15 | Max | 1.01 | OK |
| 1.70E-15 | Mean | 1.01 | OK |
| 1.30E-15 | Min | 1.01 | OK |

**6. Distance factor**

Same as 3.6 microns

**7.Neptune flux (W/m^2/um, 1 AU from Earth and Sun, from Stauffer Vega Jy)**

| | |
|---|---|
| 1.86E-09 | Max |
| 1.35E-09 | Mean |
| 1.03E-09 | Min |

**8. Solar flux at 1 AU**

E490, http://rredc.nrel.gov/solar/spectra/am0/ASTM2000.html
5.393 W/m^2/um
(Wehrli gives 0.005 W/m^2/nm at 4575.0 nm)

**9. Luminosity ratio (Neptune to Sun at 1 AU)**

| | |
|---|---|
| 3.45E-10 | Max |
| 2.51E-10 | Mean |
| 1.91E-10 | Min |

**10. Neptune size factor**

Same as 3.6 microns

**11. Neptune albedo**

| | | | |
|---|---|---|---|
| 0.0127 | Max | 1.3 | % |
| 0.0093 | Mean | 0.9 | % |
| 0.0070 | Min | 0.7 | % |

**Statistics of the observations below**

| | BMJD | Mag | Mag uncert. |
|---|---|---|---|
| Mean | 57441.16646 | 10.4770 | 0.0028 |
| St. Dev. | 0.20976 | 0.1638 | 0.0014 |
| SD Mean | 0.00088 | 0.0507 | 0.0282 |
| Count | 2016 | 2016 | 2016 |
| Min | 57440.80363 | 10.1324 | 0.0022 |
| Max | 57441.52959 | 10.7736 | 0.0441 |
| Max - Min | 0.72596 | 0.6412 | 0.0419 |

[Note: In the original spreadsheet the individual 4.5 micron observations follow at this point.]